\documentclass[aps,prx,longbibliography,twocolumn,nofootinbib, superscriptaddress,floatfix]{revtex4-2}
\usepackage[justification=justified, singlelinecheck=false]{caption}
\makeatletter
\long\def\@makecaption#1#2{%
  \par\addvspace{\abovecaptionskip}%
  \begingroup
    \small
    \setlength{\parindent}{0pt}%
    \setlength{\leftskip}{0pt}%
    \setlength{\rightskip}{0pt}%
    \setlength{\parfillskip}{0pt plus 1fil}%
    #1.\ #2\par
  \endgroup
  \addvspace{\belowcaptionskip}%
}
\makeatother
\usepackage{amsmath}
\usepackage{amssymb}
\usepackage{subfig}
\usepackage{graphicx}
\usepackage{comment}
\usepackage{soul,xcolor}
\usepackage{bbold}
\usepackage[colorlinks=true, urlcolor=blue,urlbordercolor={0 1 1}]{hyperref}
\usepackage{tikz}
\usepackage{braket}
\usepackage[english]{babel}
\usepackage{graphicx}
\usepackage{lipsum}
\usepackage{xr}
\usepackage{ulem}
\usepackage{comment}

\usepackage[version=4]{mhchem}

\newcommand{\<}{\langle}

\newcommand{\up}{\uparrow}
\newcommand{\down}{\downarrow}
\renewcommand{\>}{\rangle}
\renewcommand{\(}{\left(}
\renewcommand{\)}{\right)}

\renewcommand{\v}[1]{\mathbf{#1}} 

\renewcommand{\d}{\partial}

\newcommand{\ba}{\begin{align}}
\newcommand{\ea}{\end{align}}

\newcommand{\tb}[1]{ \textcolor{blue} }

\renewcommand{\d}{^{\dagger}}

\begin{document}

\title{Dirac node pinning from Dzyaloshinskii-Moriya interactions in a Kagome spin liquid}

\author{Ajesh Kumar}
\affiliation{Department of Physics, Massachusetts Institute of Technology, Cambridge, Massachusetts 02139, USA}

\author{Byungmin Kang}
\affiliation{Department of Physics, Massachusetts Institute of Technology, Cambridge, Massachusetts 02139, USA}

\author{Patrick A. Lee}
\affiliation{Department of Physics, Massachusetts Institute of Technology, Cambridge, Massachusetts 02139, USA}

\date{\today}
\begin{abstract}
Recent experiments on the Kagome spin liquid candidate material YCu${}_3$(OH)${}_6$Br${}_2$[Br${}_{1-y}$(OH)${}_y$] suggest the presence of Dirac fermionic spinons near the magnetization plateau at 1/9. Theories suggest that the spinons are charge neutral spin-$1/2$ excitations, in a $2\pi/3$ flux which  triples the unit cell. Generally a gap is expected, and   there is no symmetry protection for the Dirac nodes in this system. The question arises as to what causes the nodes and stabilizes them. In this work, we propose a node-creation and node-pinning mechanism driven by the Dzyaloshinskii-Moriya (DM) interactions. Employing Gutzwiller-projected variational Monte Carlo calculations, we demonstrate that DM interactions induce a band closing phase transition in the spinon spectrum. There is a change in the Chern number when the bands are inverted.  Together with the DM-generated internal gauge flux, the coupling to the spinon orbital magnetization counteracts the band reopening. This interplay energetically pins the Dirac nodes over a range of parameters, resulting in a pinning mechanism distinct from the usual one from symmetry protection.
\end{abstract}
\maketitle

\section{Introduction} 
Quantum spin liquids (QSLs) are exotic states of matter occuring in spin systems characterized by fractionalized excitations coupled to emergent gauge fields~\cite{savary2016quantum,zhou2017,broholm2020quantum}. They are expected to emerge in geometrically frustrated systems, such as the Kagome lattice, where conventional magnetic order is suppressed. There have been extensive experimental efforts to identify materials that realize a QSL state, with a major milestone being the discovery of herbertsmithite—a Mott insulator on the Kagome lattice exhibiting signatures of a QSL~\cite{norman2016,han2012fractionalized,pilon2013,fu2015evidence}. 
Recent experiments on YCu${}_3$(OH)${}_6$Br${}_2$[Br${}_{1-y}$(OH)${}_y$] (YCOB), a relative of herbertsmithite, have reported  signatures suggestive of a QSL, including quantum oscillations in the magnetic susceptibility, a 
$T^2$ specific heat at low temperatures, and a $1/9$ magnetization plateau~\cite{zheng2025unconventional,zheng2024thermodynamicevidencefermionicbehavior,jeon2024one,suetsugu2024}. These observations were interpreted with a scenario involving Dirac spinons in a tripled unit cell \cite{zheng2025unconventional}.
The $1/9$ plateau was predicted theoretically and interpreted in terms of a $\mathbb{Z}_3$ spin liquid~\cite{nishimoto2013controlling} or a valence bond solid phase~\cite{picot2016,fang2023}. 
A recent  variational Monte Carlo study based on the $2\pi/3$ flux per unit cell~\cite{he2024} found a ground state energy very close to that given by DMRG and consistent with the $\mathbb{Z}_3$ spin liquid interpretation.  However, these states generally host gapped spinons, and if Dirac nodes do appear, they require fine-tuning and are unstable under small perturbations. While symmetry protection can stabilize Dirac nodes—as in graphene, where Dirac nodes are protected by time-reversal and inversion symmetries —this raises a key question: what mechanism stabilizes them in the case of YCOB, where time-reversal symmetry is strongly broken due to an applied magnetic field?

In this Letter, we investigate how Dirac nodes can be created and then  stabilized in the spinon spectrum of the $2\pi/3$-flux state. Since they lack symmetry protection, Dirac nodes arise at fine-tuned critical points of band inversion transitions, prompting the question of 
whether there exist mechanisms that maintain the system at this critical point. We identify two opposing tendencies at play which compete and result in energetically stabilizing the Dirac nodes. Our starting point will be the gapped state stabilized in previous variational Monte Carlo (VMC) calculations~\cite{he2024}. 
We will first address the tendency of the system to try to undergo a band inversion transition. Using VMC calculations we will demonstrate that Dzyaloshinskii-Moriya (DM) interactions, whose strength is denoted by $D$, tend to bring down the spinon gap and eventually go through band inversion at a critical strength $D_c$. 

The second part of our argument addresses the opposing mechanism that resists band reopening after the band inversion transition. Here, an internal orbital magnetic field coupling to the spinons plays a crucial role. In YCOB, the presence of such a field is suggested by the observed quantum oscillations.
The applied physical magnetic field can induce an out-of-plane gauge magnetic field $b$, either via coupling to the spin-chirality term~\cite{motrunich2006orbital} or, through DM interactions which is likely more relevant to YCOB~\cite{kang2024}. This field should be regarded as a perturbation on top of the existing $2\pi/3$ flux.
Importantly, there is a jump in the Chern number at the band inversion. By analyzing the change  of the orbital magnetization of the spinon bands \cite{xiao2005, ceresoli2006} at the band inversion transition, 
we demonstrate that the resulting change in orbital magnetization, $\delta M$, induces an energy change $-\delta M b$, which when positive, suppresses band inversion and energetically stabilizes the Dirac nodes. 

We now detail the first part of our argument that shows that DM interactions can drive the band inversion transition. 

\section{DM-driven band inversion}

\begin{figure}[t]
\includegraphics[width = 0.9\columnwidth]{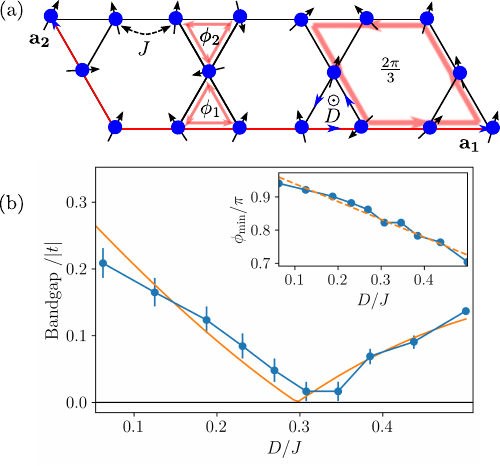}
\caption{(a) Schematic of the variational state used in our calculations. There is a $2\pi/3$ gauge flux per Kagome unit cell, which leads to a tripling of the magnetic unit cell, formed by the lattice vectors shown in blue. The state is parameterized by fluxes through the up and down triangles $\phi_1$ and $\phi_2$ respectively.
The spins indicated by the arrows are coupled via a nearest-neighbor Heisenberg interaction and a Dzyaloshinskii-Moriya (DM) interaction defined by a DM vector pointing out of the plane. The DM interaction is illustrated within one of the triangles and acts in a counter-clockwise sense, as shown by the blue arrows on the bonds. 
(b) The bandgap is obtained from the bandstructure of the mean-field Hamiltonian in Eq.~\eqref{eq:mf-ham} at the optimized flux values using Gutzwiller projected variational Monte Carlo (VMC) calculations (inset). The figure indicates a gap closing near $D_c \approx 0.28 J$. The system size used was $4 \times 6 \times 9$, where $4$ and $6$ indicate the number of magnetic unit cells along $\v a_1$ and $\v a_2$ respectively, and there are $9$ sites 
in the tripled magnetic unit cell. Our calculations show that the energy is minimized for $\phi_2=\phi_1$, and we denote the corresponding value as $\phi_{\textrm{min}}$ (see Appendix~\ref{app_numerical}).
The inset shows the optimized value of the flux $\phi_{\text{min}}$ as a function of $D$, which is well-fit by a linear curve shown as a dashed orange line: $\phi_{\text{min}}/\pi = -0.54 D/J + 0.99$. 
Using the linear fit, we obtain the corresponding bandgap for the optimized value of flux, which is plotted as the orange solid line.
}
\label{fig_vmc}
\end{figure}

We consider the following spin model on the Kagome lattice with Heisenberg and DM terms in the presence of an external Zeeman magnetic field $B$, which we set to be positive.
\begin{align}
    H = \sum_{\<i,j\>} \Big[ J\vec{S}_i \cdot \vec{S}_j + D \hat{z} \cdot \big( \vec{S}_i \times \vec{S}_j\big) \Big] + \frac{g_s \mu_B B}{\hbar} \sum_i S^z_i
\end{align}
For the DM term, we adopt an ordering convention such that the vectors joining sites $i$ and $j$ form a counter-clockwise loop on each of the triangular plaquettes. With this convention, it has been experimentally shown that $D>0$ in a closely related Kagome material where Br is replaced by Cl and is ordered~\cite{zorko2019negative}. Introducing charge-neutral fermionic spinons $f_{i\alpha}$, $\vec{S}_i = \frac{1}{2} f\d_{i\alpha} \vec{\sigma}_{\alpha \beta} f_{i\beta}$, where $\vec{\sigma}$ are spin Pauli matrices, we focus on the following variational Hamiltonian with $2\pi/3$ gauge flux.
\begin{align}
\label{eq:mf-ham}
    H_f[\phi_1,\phi_2] = -\sum_{\<i,j\>, \alpha} t_{ij}[\phi_1,\phi_2] f\d_{i\alpha}f_{j\alpha} - \sum_{i,\alpha} \mu_{\alpha}[\phi_1,\phi_2] f\d_{i\alpha}f_{i\alpha}
\end{align}
We pick the spinon hoppings $t_{ij}$ to have a uniform amplitude, however with a non-uniform phase pattern, which is parameterized by the flux through the up/down triangles $\phi_{1/2}$ respectively, as shown in Fig.~\ref{fig_vmc}. For the Heisenberg model, the state with uniform hopping amplitude has been shown to be the lowest energy state in previous VMC calculations~\cite{he2024}. Further, we have assumed that fluxes $\phi_{1/2}$ are spin-independent.
The chemical potentials $\mu_{\alpha}$ are chosen so as to obtain states with $1/9$-magnetization, i.e., five (four) $\down$ ($\up)$-spin bands occupied, and an average filling of one spinon per site. 
We constrain ourselves to the consideration of states with $1/9$-magnetization motivated by the experimental observations of this state and previous VMC calculations~\cite{he2024} that find a range of magnetic fields where this state is stablilized.
The mean-field state $|\psi_{mf}[\phi_1,\phi_2]\>$ is then projected onto the physical subspace where every site has exactly one spinon: $|\psi_{proj}[\phi_1,\phi_2]\> = P_G |\psi_{mf}[\phi_1,\phi_2]\>$. 
We calculate the energy $E[\phi_1,\phi_2] = \<\psi_{proj}[\phi_1,\phi_2]| H|\psi_{proj}[\phi_1,\phi_2]\>$ by a Monte Carlo sampling~\cite{gros1989physics}, which is then minimized over $\phi_{1/2}$ to obtain the ground state. In the Monte Carlo sampling, we use a skip of $100$ samples, which ensures that successive configurations used for sampling are weakly-correlated.
In our calculations, we find that the energy is minimized for $\phi_2=\phi_1$ (see Appendix~\ref{app_numerical}), therefore, we present results as a function of $\phi_1$ only. Fig.~\ref{fig_vmc} (b) (inset) shows the optimized value of $\phi_1$, defined as $\phi_{\text{min}}$, as a function of $D$. Without the DM term, the optimized value is $\phi_{\text{min}} \approx \pi$. 
The corresponding bandgap between the fifth and sixth bands in the $\down$-spin sector was calculated using the mean-field Hamiltonian $H_f$, in units of $|t|$. The variational wavefunction and hence the variational energy do not depend on $|t|$, as it only sets the overall scale of the Hamiltonian. Typically, we find $|t| \approx 0.26J$, in our mean-field calculations discussed in Sec.~\ref{sec_energetics}. 

We note that the $2\pi/3$-flux state is two-fold degenerate. Acting by the time-reversal operation results in a state with $-2\pi/3$ flux per unit cell and with the spins exchanged. Importantly, the single-particle states in both the spin sectors are the same. Consider the state with $-2\pi/3$ flux but the same spin magnetization. Since the Heisenberg and DM interactions are time-reversal invariant and the Zeeman energy remains unchanged, this state retains the same energy as the original one.

\begin{figure*}[t]
\centering
\includegraphics[width = 0.8\textwidth]{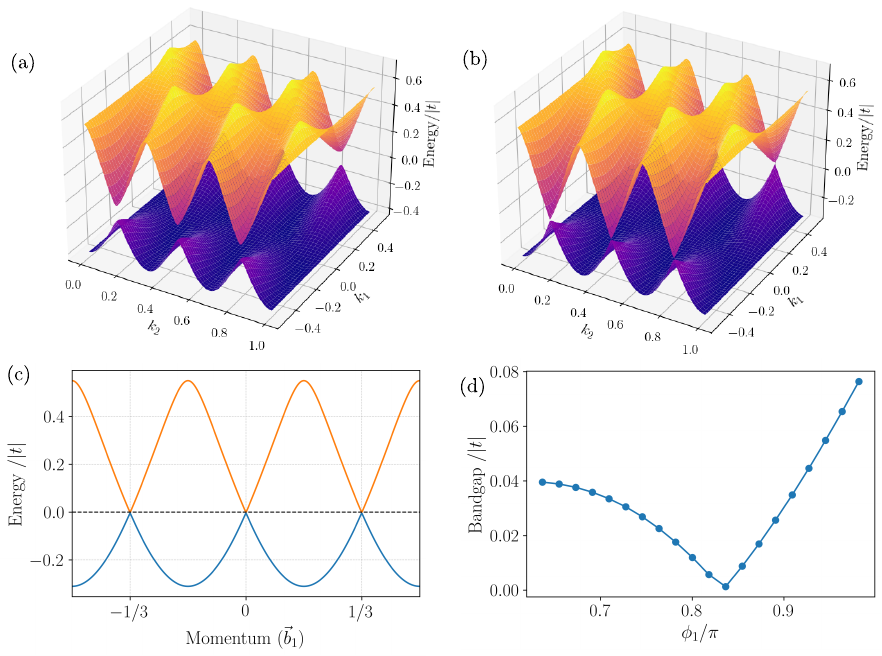}
\caption{(a, b) Bands 5 and 6 plotted in the magnetic Brillouin zone for $D = D_c/2$ and $D=D_c$ respectively, showing the closing of the bandgap near three Dirac nodes.
(c) The dispersion in units of $|t|$ of spin-$\down$ bands $5$ and $6$ along the reciprocal lattice vector $\v b_1 = 2\pi \hat{y}/(\sqrt{3}a)$, where $a$ is the inter-site distance, for $D = D_c$, showing three equally-spaced Dirac crossings. Dashed lines indicate the chemical potential. 
(d) The bandgap between bands 5 and 6 computed for the variational mean-field state $|\psi_{mf}\>$ shows a gap closing at $\phi_1 \approx 0.84 \pi$. The horizontal axis is the flux that minimizes the ground state energy for a given $D$. Its relationship with $D$ is given in the inset of Fig.~\ref{fig_vmc} (b). Note that $D$ is increasing for decreasing value of $\phi_{\text{min}}$.
}
\label{fig_bands}
\end{figure*}

In Fig.~\ref{fig_vmc} (b), we compute the bandgap from the bandstructure of the mean-field Hamiltonian in Eq.~\eqref{eq:mf-ham} at the optimized flux values obtained from the VMC computation (inset). As shown in the figure, the bandgap closes for $D =D_c \approx 0.28 J$. 
The closing of the bandgap is robust across different system sizes, as shown in Appendix~\ref{app_numerical}, and is reproduced in the mean-field analysis presented in Sec.~\ref{sec_energetics}.
At this value of $D$, the mean-field  bandstructure, plotted in Fig.~\ref{fig_bands} (c), consists of three Dirac nodes corresponding to a change in the Chern number of the fifth spin-$\down$ band from $-2$ to $1$. Fig.~\ref{fig_bands} (d) shows the bandgap calculated using $H_f$, as a function of $\phi_1$. The bandgap closes at $\phi_1=\phi^c_1 \approx 0.84 \pi$, consistent with the bandgap closing demonstrated in Fig.~\ref{fig_vmc} (b).

Using our VMC calculations, we have thus demonstrated that DM interactions can drive the system through a band inversion transition. Next, we investigate the orbital magnetism of the spinon bands, which, as we will show, can lead to pinning the system at the band inversion transition, leading to Dirac nodes over a range of $D$.

\section{Orbital magnetism }

\begin{figure}[t]
\includegraphics[width = 0.88\columnwidth]{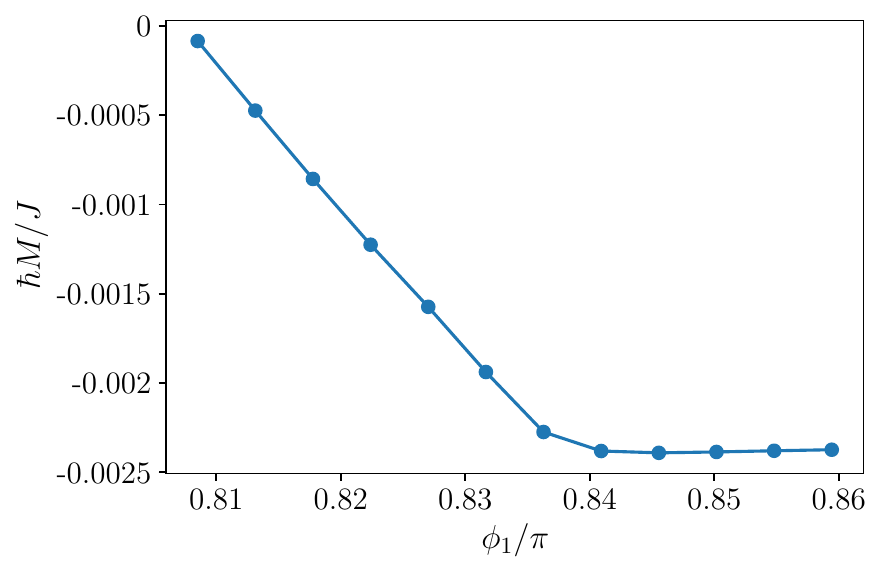}
\caption{Orbital magnetization computed numerically for the variational state $|\psi_{mf}\>$ versus the optimized value of $\phi_1$. 
}
\label{fig_om}
\end{figure}

We have so far discussed states having a $2\pi/3$ flux per unit cell. As is typical for QSL states, this flux is associated with an internal gauge field. The fractionalization of spins in terms of spinons, $\vec{S}_i = \frac{1}{2} f\d_{i\alpha} \vec{\sigma}_{\alpha \beta} f_{i\beta}$, introduces a $U(1)$ gauge redundancy~\cite{wen2004quantum}. Namely, the phase of the fermionic operators can be changed locally without affecting the physical spin operators. A complete low-energy description therefore involves an emergent fluctuating $U(1)$ gauge field $\v a$ that couples minimally to $f$. The fluxes discussed in this work correspond to the mean gauge flux of $a$. We denote the associated internal magnetic field by $b=\v{\nabla} \times \v{a}$ and its flux per unit cell by $\Phi_b$ (both measured relative to $2\pi/3$ flux). 
It is important to emphasize that the spinons, being charge neutral, do not couple orbitally to an external magnetic field, responding only through the Zeeman term. 

We now address how the mean value of $\Phi_b$ is modified in the presence of DM interactions. The physical meaning of the gauge flux $\Phi_b$ is the scalar spin chirality $\chi$  =
$\langle \vec{S}_1 \cdot \vec{S}_2 \times \vec{S}_3 \rangle$ of three neighboring spins~\cite{wen1989chiral}. In the presence of DM, $\langle \vec{S}_2 \times \vec{S}_3 \rangle \neq 0$. Therefore the factorized form of $\chi \approx \langle S_z \rangle \langle \vec{S}_2 \times \vec{S}_3 \rangle_z$ is finite in the presence of spin-polarization $\langle S_z \rangle \neq 0$. A calculation beyond the factorization approximation was done in Ref.~\cite{kang2024} showing that the DM term plus spin magnetization contributes a spinon Berry phase in each Kagome unit cell. Thereby, on top of the $2\pi/3$ flux, the DM term generates an additional flux $\Phi_b$. 
In this section, we will study the role of this flux. 
In Ref.~\cite{kang2024}, $\Phi_b$, was found to be: $\Phi_b = 7.2D \< \sigma_z \>/J$. Using the YCOB unit cell area $38.53$~\AA$^2$ and $\< \sigma_z \> = -1/9$ at the magnetization plateau, we find $\Phi_b = 600$ T for the maximum value of $D$ used in this work: $D=J/2$. 
This value is much smaller than the orbital field of $\mathcal{O}(10^4)$ T that is responsible for the $2\pi/3$ flux. 
Hence, we will treat $b$, the internal magnetic field coupled to spinons, as a perturbation on the $2\pi/3$-flux state. 
The leading linear order correction to the energy then arises due to the orbital magnetization of the spinons, which are characterized by their own band structure. We emphasize that this orbital magnetization is conjugate to the internal gauge magnetic field $b$, and not the physical external magnetic field.
The $b$-field also splits the two-fold degenerate $2\pi/3$-flux states since they carry opposite orbital magnetization. 

The orbital magnetization contributed by spinon band $n$, $M_n$, is given by~\cite{xiao2005,ceresoli2006,zhu2020,kang2025orbitalmagnetizationmagneticsusceptibility,liu2025orbitalmagnetizationcorrelatedstates}
\begin{align}
    \hbar M_n = \text{Im} \int \frac{d^2 \v k}{(2\pi)^2} \sum_{n' \neq n} &\frac{\<n|\partial_x H_f|n'\>\<n'|\partial_y H_f|n\>}{\(\epsilon_n(\v k)-\epsilon_{n'}(\v k)\)^2} \times \nonumber\\
    & \big(\epsilon_n(\v k) + \epsilon_{n'}(\v k) - 2\mu\big)
\end{align}
where $n'$ is summed over bands, and $|n\>$, $|n'\>$ denote Bloch states at momentum $\v k$. Summed over all the occupied states in both spin sectors, we obtain the total orbital magnetization $M$. 
In Fig.~\ref{fig_om}, we plot $\hbar M$ as a function of $\phi_{\text{min}}$, where the $\phi_{\textrm{min}}$ is obtained from the VMC calculations and the bands are obtained from the mean-field Hamiltonian Eq.~\eqref{eq:mf-ham} at the optimized flux $\phi_{\textrm{min}}$. 
It is important that the chemical potential is set to be at the bottom of the sixth spin-$\down$ band for $D>D_c$ and at the top of the fifth spin-$\down$ band for $D < D_c$, to ensure the half-filling condition. See Appendix for an explanation of this using a low-energy Dirac model. We find a linear increase in $M$ as a function of $\delta \phi_1 = \phi_1-\phi_1^c$. 
From the inset of Fig.~\ref{fig_vmc} (b), where we show that the variation of $\phi_{\text{min}}$ versus $D$ is well-fit by a linear curve, we see that there is a linear increase in $M$ versus $D$ for $D>D_c$.

The energy contribution, due to the orbital magnetism, per unit cell can be expressed as $E_M = -\hbar M \Phi_b$, where $\Phi_b$ is the flux of $b$ through a unit cell. 
Since the $b$-field generated by DM interactions was shown to be opposite to the applied magnetic field~\cite{kang2024}, and therefore, negative in our sign convention, $E_M$ is positive and corresponds to an energy penalty for the reopening of the bandgap. 
The question remains, however, whether this effect is strong enough to counter the tendency to reopen the gap that was explained in the previous section. We now address this question through a mean-field analysis.
It would be very interesting to answer this question using Gutzwiller-projected VMC calculations by considering variational wavefunctions that incorporate the incommensurate fluxes $\Phi_b$. The next step would be to find the optimal $\Phi_b$ that minimizes the energy. However, working with small system sizes one is constrained to a small discrete number of fluxes $\Phi_b$ since they have to be commensurate with the total system area. We leave such investigations to future work. 

\section{Energetic competition and pinning of the Dirac node}
\label{sec_energetics}
\begin{figure}[t]
\includegraphics[width = 0.88\columnwidth]{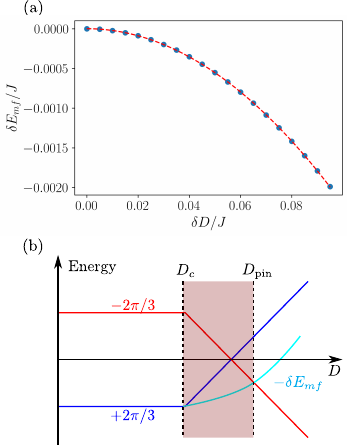}
\caption{(a) The mean-field energy (blue points) $\delta E_{mf}$ of the optimal state relative to the Dirac state, in the absence of a gauge magnetic field. The red curve is a quadratic fit: $\delta E_{mf} \approx 0.22 \delta D^2/J$.
(b) Schematic of the energy competition. The blue/red curve denotes evolution of the energy contributed by the orbital magnetization $E_M$ of the $\pm 2\pi/3$-flux states. When the magnetization crosses zero, the ground state switches from $+2\pi/3$ flux to to $-2\pi/3$-flux.
The $-\delta E_{mf}$ curve is shown in cyan, which crosses the $E_M$ curve at $D=D_{\text{pin}}$. For $D_c < D < D_{\text{pin}}$, the Dirac state is energetically pinned to be the ground state. The Chern number changes from $-2$ for $D<D_c$ to $-1$ for $D>D_{\text{pin}}$. 
}
\label{fig_energetics}
\end{figure}
We study the mean-field energy: $E_{mf}[\phi_1,\phi_2] = \<\psi_{mf}[\phi_1,\phi_2]|H|\psi_{mf}[\phi_1,\phi_2]\>$ as a function of $\phi_1=\phi_2$, as before. Defining $\chi_{ij,\alpha} = \< f\d_{i\alpha} f_{j \alpha}\>$, the mean-field energy can be expressed as: 
\begin{align}
    E_{mf} &= \sum_{\langle ij \rangle}  J \langle \vec{S}_i \rangle \cdot \langle \vec{S}_j \rangle - D\text{Im} (\chi_{ij,\downarrow} \chi_{ij,\uparrow}^*)  \nonumber \\
    &\quad - \frac{J}{4} \left( |\chi_{ij,\uparrow}|^2 + |\chi_{ij,\downarrow}|^2 + 4 \, \text{Re}(\chi_{ij,\uparrow} \chi_{ij,\downarrow}^*) \right).
\end{align}
We determine the optimal value of $\phi_1$ at which $E_{mf}$ is minimized, denoted as $\phi_{\text{min}}$, and find that within this mean-field treatment, $D_c \approx 0$.

To analyze the stability of the Dirac state, we compute the mean-field energy at $\phi_{\text{min}}$ relative to the mean-field energy of the Dirac state $\delta E_{mf}$, as a function of $D$. 
Our results are presented in Fig.~\ref{fig_energetics}(a), which show that 
\begin{align}
\delta E_{mf} \approx -0.22 (\delta D)^2/J, 
\end{align}
where $\delta D = D-D_c$. 

We also find the orbital magnetization at $\phi_{\text{min}}$ relative to the Dirac state, from the data shown in Fig.~\ref{fig_om}: $\delta M \approx 0.25 \delta D/\hbar$, for $\delta D>0$. 
According to Ref.~\cite{kang2024}, 
the emergent gauge flux $\Phi_b$ is proportional to the magnetization and the ratio $D/J$. Near the 1/9 plateau we use the value $\Phi_b = -\kappa D/J$, with an estimated proportionality constant $\kappa =0.8$. 
This leads to an energy penalty for gap opening, given by 
\begin{align}
    \delta E_M \approx 0.25 \kappa \delta D \(\delta D + D_c\)/J.
\end{align}
We note that the sign of $\kappa$ is crucial. If $\kappa$ has the opposite sign, this term and the linear parts in Fig. 4(b) will have opposite sign, resulting in an energy gain rather than a penalty. In that case the gap will re-open without any pinning at the Dirac point.
With the current sign, for small $\delta D$, this linear penalty will dominate over the quadratic-in-$\delta D$ energy gain $\delta E_{mf}$, resulting in the stabilization of the Dirac state for a range of $D$. 

To obtain an estimate for this range, it is important to consider the evolution of the orbital magnetization of the ground state. The magnetization of the $2\pi/3$-flux state plotted in Fig.~\ref{fig_om} approaches zero, where it intersects the magnetization curve of the $-2\pi/3$-flux state and eventually becomes positive. Correspondingly, the magnetization energy $E_M$ curves of the two states cross, resulting in a switch of the ground state from the $+2\pi/3$-flux state to the $-2\pi/3$-flux state. 
This is schematically illustrated in Fig.~\ref{fig_energetics}(b). The pinning of the Dirac state survives up to $D=D_{\text{pin}}$, where the energy gain due to gap opening $\delta E_{mf}$ overwhelms the energy penalty due to $E_M$. Using the mean-field parameters, we estimate the width of the region where the Dirac state is pinned to be $\delta D_{\text{pin}}\equiv D_{\text{pin}}-D_c \approx 0.056 D_c$. We propose that the experimental system may have a DM term with the magnitude $D$ that places it in the pinned Dirac state. 

Let us now discuss the nature of the transition at $D=D_{\text{pin}}$. In the region where the Dirac node is pinned, the state corresponds to $\phi_{\text{min}}=\phi_1^c$, across the entire window. Let us define the value of $\phi_{\text{min}}$ at $D=D_{\text{pin}}$, in the absence of the $b$-field as $\phi^{\text{pin}}_1$. At $D=D_{\text{pin}}$, the ground state jumps from $\phi_1^c$ to $\phi_1^{\text{pin}}$ via a first-order phase transition. Consequently, the bandgap also exhibits a discontinuous jump at this transition.

The switch in the sign of the flux has a rather non-trivial consequence for the change in the Chern number as the gap reopens. For the $2\pi/3$-flux state, the Chern number of the fifth band changes from $C=-2$ to $1$ at the band inversion transition corresponding to three Dirac nodes of the same chirality. However, switching to the $-2\pi/3$-flux state results in a band with $C=-1$, instead of $+1$.


\section{Discussion}
Using VMC and parton mean-field calculations, we have demonstrated an energetic mechanism for stabilizing a Dirac state of spinons, in the absence of any symmetry protection. Our results highlight the crucial role of Dzyaloshinskii-Moriya interactions and the orbital magnetism of the spinon bands in pinning the Dirac nodes and preventing gap reopening. Although we have presented supporting numerical results based on VMC and parton mean-field calculations, we emphasize that our proposed mechanism does not rely on parameter fine-tuning or other numerically sensitive details.

Our pinning mechanism is different from previous examples which rely on symmetry protection and may have broader applicability beyond spin liquid states. As an example, we discuss potential extensions to electronic systems. In this work, the spinon bands exhibited orbital magnetism due to the presence of a $2\pi/3$ background gauge flux. Recent experiments on two-dimensional materials have reported orbital magnetism arising from spontaneous valley polarization of electrons, which breaks time-reversal symmetry~\cite{sharpe2019emergent,serlin2020intrinsic,han2024correlated}. Moreover, displacement-field-driven transitions between states with zero and finite Chern numbers have been observed~\cite{chen2020tunable}. It would be intriguing to investigate whether an out-of-plane orbital magnetic field could stabilize a critical state over a finite range of displacement fields in valley-polarized systems, providing an analog to the pinning mechanism explored in this study. 

\section{Acknowledgements}
The authors acknowledge the MIT SuperCloud and Lincoln Laboratory Supercomputing Center for providing HPC resources that have contributed to the research results reported within this manuscript.
A.K. was supported by the Gordon and Betty Moore Foundation EPiQS Initiative through Grant No. GBMF8684 at the Massachusetts Institute of Technology. P.A.L. acknowledges support from DOE (USA) office of Basic Sciences Grant No. DE-FG02-03ER46076.

\bibliography{references.bib}

\appendix

\onecolumngrid
\newpage

\section{Additional numerical results}
\label{app_numerical}

\begin{figure}[h]
    \centering
    \subfloat[Heisenberg term ($J=1$).]{\includegraphics[width=0.45\textwidth]{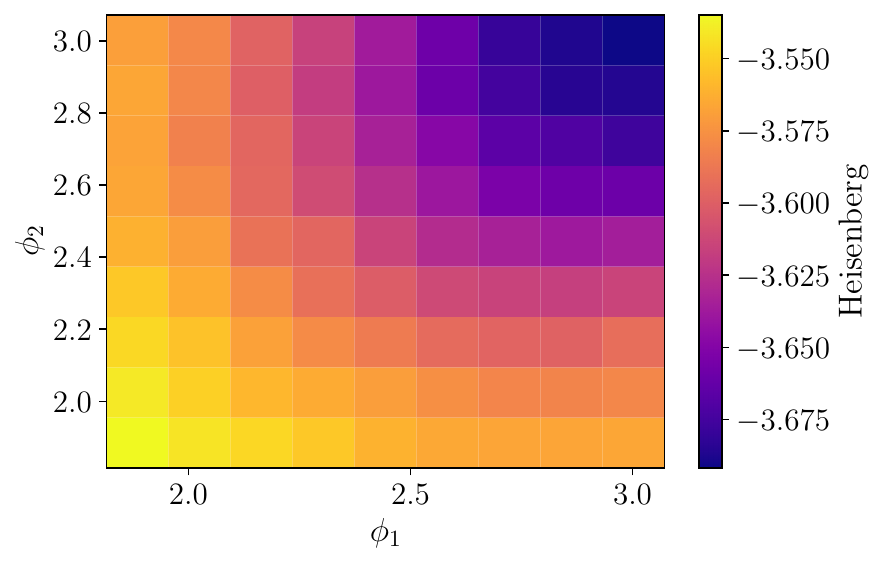}}
    \hfill
    \subfloat[DM term ($D=1$).]{\includegraphics[width=0.45\textwidth]{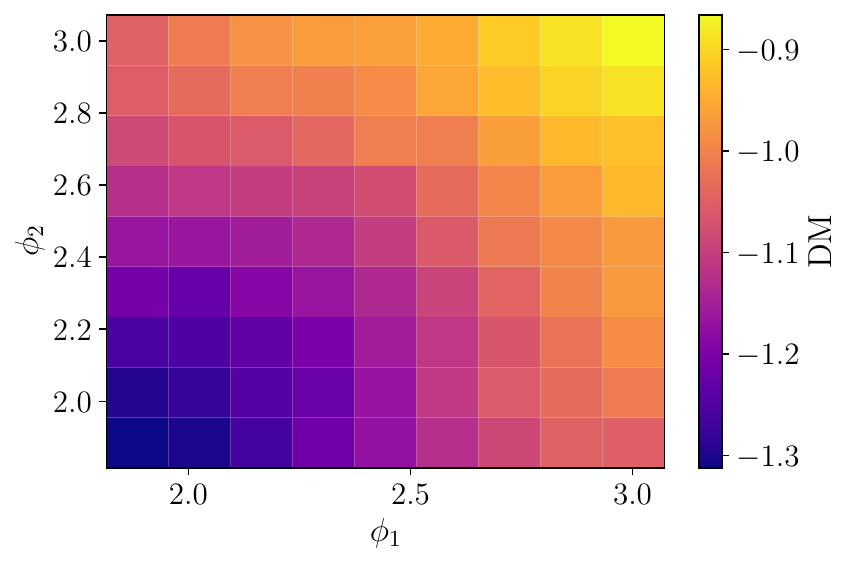}}
    \caption{Energies evaluated in VMC as a function of fluxes $\phi_{1/2}$ for system size $1 \times 6 \times 9$ to demonstrate that energy minima lie on the $\phi_1=\phi_2$ line.}
    \label{fig:subfigures}
\end{figure}

\begin{figure}[h]
\includegraphics[width = 0.5\columnwidth]{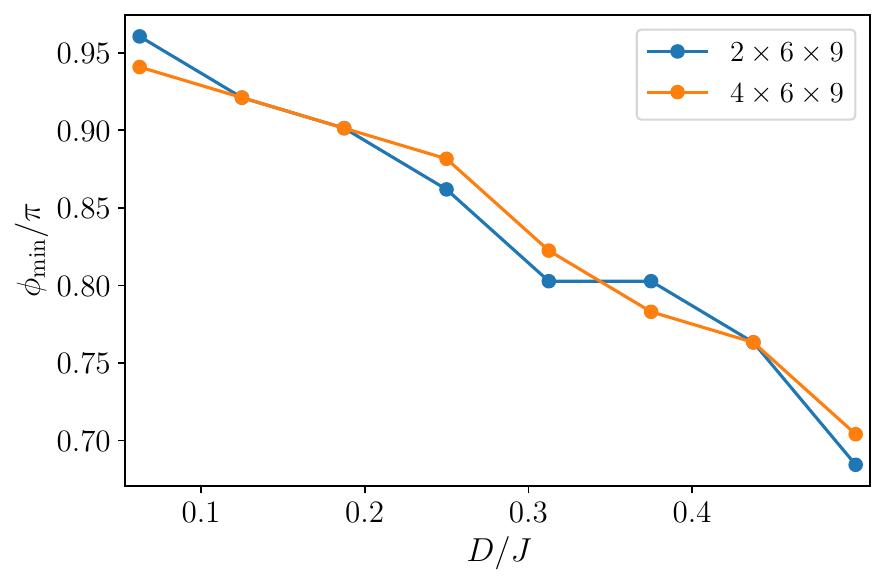}
\caption{Comparison of the optimized value $\phi_{\text{min}}$ obtained as a function of $D/J$ for different system sizes. Importantly, in both cases, $\phi_{\text{min}}$ crosses the value $\phi^c_1 \approx 0.84 \pi$ where the bandgap closes. 
}
\label{fig_syssize}
\end{figure}

\section{Orbital Magnetism in Dirac model}
\label{app_dirac}

\begin{figure}[t]
\includegraphics[width = 0.5\columnwidth]{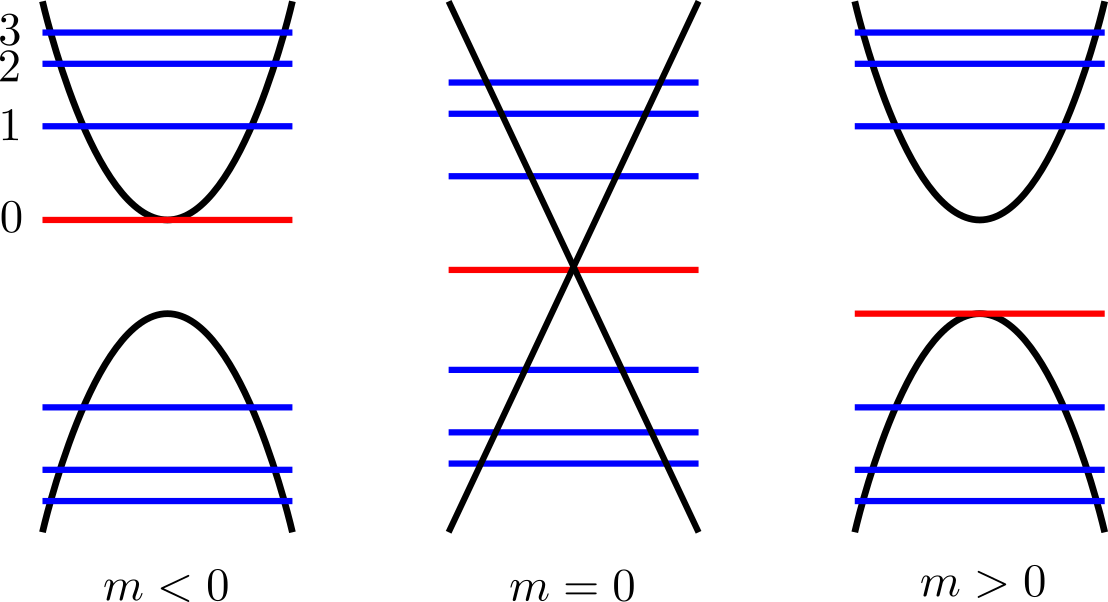}
\caption{Schematic showing the Dirac Landau levels across the band inversion transition. The two bands (in the absence of magnetic field) are shown in black. Landau levels with $n>0$ are shown in blue. The anomalous Landau level at $n=0$ is shown in red. Notably, it moves from the bottom of the conduction band for $m<0$ to top of the valence band for $m>0$. The chemical potential follows the anomalous Landau level to satisfy the half-filling condition.
}
\label{fig_diracLL}
\end{figure}

In this section, we describe the orbital magnetism near the band-inversion transition using a simple low-energy model, which will serve well to illustrate qualitatively the linear increase in the magnetization found in the main text. Our discussion is along the lines of Ref. \cite{koshino2010,dong2024}. Consider the massive Dirac Hamiltonian 
\begin{align}
    H_D = \begin{pmatrix}
        &m &v (k_x + ik_y) \\
        &v (k_x - ik_y) &-m
    \end{pmatrix}
\end{align}
The Chern number of the lower band changes by $+1$ when the system goes through a band inversion transition, i.e., when the mass $m$ changes sign from positive to negative~\cite{bernevig2013topological}. We now calculate the orbital magnetization by introducing a magnetic field $\v b = \nabla \times \v a$, that couples to the orbital motion of the spinons. Introducing the momentum $\v \Pi = \v k - \v a$ we solve for the Landau levels in terms of the ladder operators $d = \(\Pi_x + i\Pi_y\)l_b/\sqrt{2}$, where $l_b$ is the magnetic length associated with $\v b$. We obtain the Landau level spectrum: $\epsilon_{n,\pm} = \pm \sqrt{\hbar^2\omega_b^2 n + m^2}$ for positive integers $n$, where $\omega_b = \sqrt{2}v/l_b$. Additionally, there is an anomalous Landau level at $\epsilon_{n=0} = -m$. 

The spinon chemical potential $\mu$ must be at $\mu = \epsilon_0$ to satisfy the half-filling condition on average for the spinons. Crucially, this implies that $\mu=-m$ is at the top of the valence band for $m>0$ and moves to the bottom of the conduction band for $m<0$. The orbital magnetization $M$ (in the out-of-plane direction) is defined in terms of the thermodynamic potential $\Omega$ as $M = -\partial \Omega/\partial b$~\cite{koshino2010,dong2024}. The thermodynamic potential is given by $\Omega = \sum_{n,\pm} \(\epsilon_{n,\pm}-\mu\) f(\epsilon_{n,\pm}) + \(\epsilon_{0}-\mu\) f(\epsilon_{0})$, where $f$ denotes the Fermi-Dirac distribution. The anomalous Landau level does not contribute to $\Omega$ because of the half-filling condition explained above. We calculate the orbital magnetization to find:
\begin{align}
    M &= -\frac{m}{hc}, \quad \text{for } m < 0, \nonumber \\
    M &= 0, \quad \text{for } m > 0.
    \label{eq_om_dirac}
\end{align}
consistent with the linear increase in $M$ found in Fig.~3 of the main text. 



\end{document}